# Predicting T-Cell Receptor Specificity

Tengyao Tu
*Department of Engineering Mathematics*
University of Bristol
*Bristol, United Kingdom*
vs22203@bristol.ac.uk

Wei Zeng
*Department of Engineering Mathematics*
University of Bristol
*Bristol, United Kingdom*
uo23026@bristol.ac.uk

Kun Zhao
*Department of Engineering Mathematics*
University of Bristol
*Bristol, United Kingdom*
un23692@bristol.ac.uk

Zhenyu Zhang
*Department of Engineering Mathematics*
University of Bristol
*Bristol, United Kingdom*
uj23672@bristol.ac.uk

*Abstract*—Researching the specificity of TCR contributes to the development of immunotherapy and provides new opportunities and strategies for personalized cancer immunotherapy. Therefore, we established a TCR generative specificity detection framework consisting of an antigen selector and a TCR classifier based on the Random Forest algorithm, aiming to efficiently screen out TCRs and target antigens and achieve TCR specificity prediction. Furthermore, we used the k-fold validation method to compare the performance of our model with ordinary deep learning methods. The result proves that adding a classifier to the model based on the random forest algorithm is very effective, and our model generally outperforms ordinary deep learning methods. Moreover, we put forward feasible optimization suggestions for the shortcomings and challenges of our model found during model implementation.

## I. INTRODUCTION

Cancer is the leading cause of death worldwide. In 2022, there were an estimated 20 million new cancer cases and 9.7 million deaths [1]. Although traditional methods such as surgery and chemotherapy are effective in treating cancer to a certain extent, they are often not effective in preventing the metastatic spread of the disease through disseminated tumor cells [2]. With the deepening research on the immune system, immunotherapy that can harness the immune system to fight cancer has firmly established itself as a novel pillar of cancer care [3].

The specificity of T cell receptor (TCR) plays a crucial role in immunotherapy. As an essential class of lymphocytes in the immune system, T cells can recognize and bind to antigen peptides from the major histocompatibility complex (MHC) through the TCR on the surface to trigger an immune response to recognize and attack pathogens and abnormal cells. Researching the specificity of TCR contributes to the development of immunotherapy. Still, the various experimental methods used to identify the interactions between TCRs and peptides presented by MHC molecules (pMHCs) have limitations, such as being time-consuming, costly, or technically demanding [4]. Therefore, algorithms that can accurately identify and predict TCR specificity need to be developed.

We established a TCR generative specificity detection framework consisting of an antigen selector and a TCR classifier based on the Random Forest algorithm, which can efficiently predict TCR specificity.

## II. LITERATURE REVIEW

We researched various computational models developed for TCR specificity prediction; then, these models were divided into three categories according to the working principle.

The first class of models predicts the specificity of TCR based on three-dimensional structural information of TCR and pMHC. These models perform well when there is a need for detailed knowledge and high-resolution prediction of binding patterns and specificity. However, the structural basis of TCR activation is poorly understood [5], which reduces the accuracy and feasibility of the model.

The second class of models is based on the sequence information of the TCR and pMHC. Developing high-throughput sequencing methods and single-cell RNA sequencing technologies has enabled efficient and rapid derivation of large numbers of TCR sequences from donor samples, which are collated into rich datasets [6]. It promotes the application of deep learning methods in the modeling of TCR-pMHC interactions. However, these models only utilize the sequence information of the TCR and ignore structural details. Therefore, the accuracy of this model class may not be ideal if structural features need to be considered in the prediction.

The third class of models combines the characteristics of the first two models, which can make full use of structure and sequence information to improve accuracy and reliability. In other words, it also has higher requirements for the variety and quantity of data. Therefore, these models may cause high computational costs and training difficulties if computing resources and training data are limited.

## III. METHODOLOGY

The framework of our model is depicted in Figure 1. The model includes two key components: (1) antigen selector. It is used to calculate and edit the distance matrix of TRBs between the target TCR and other TCRs in the training database. Then, the antigen of the closest training data is selected as the target antigen. (2) TCR classifier based on the Random Forest algorithm. First, the chosen target antigen and CDR3 are encoded by N-gram. Then, the encoded CDR3 and antigen sequences are used as feature input, and the labels are used as a model output to build a random forest model, which finally produces prediction results. In the following section, we will provide detailed explanations of our model.

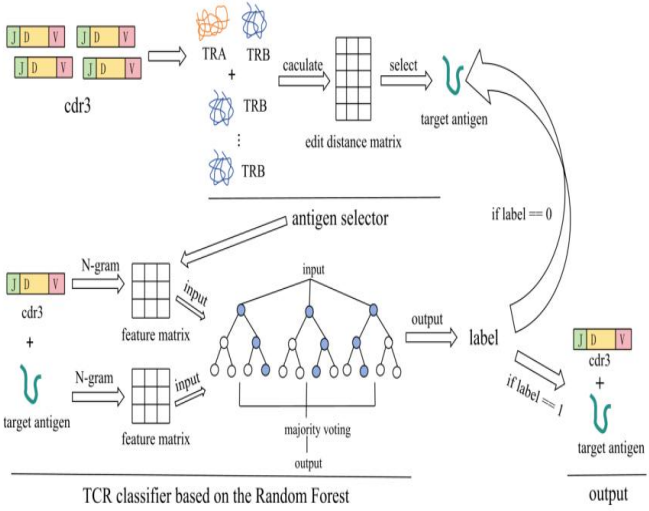

Fig.1 Model framework

Edit distance: Vujovic M [7] used the distance matrix of TCR for clustering in their paper. Inspired by this, the first part of our model framework also selects candidate antigens through the distance matrix of TCR.

When calculating the distance between sequences, we consider the edit distance, the minimum number of times a sequence can transfer into another sequence by adding, deleting, and modifying characters. For each change in the sequence, three cases are considered. Through calculation, we can get the formula for editing distance as follows:

$$D[i,j] = D[i-1,j] + 1 \; if \; the \; next \; step \; is \; deleting \quad (1)$$

$$D[i,j] = D[i,j-1] + 1 \; if \; the \; next \; step \; is \; adding \quad (2)$$

$$D[i,j] = D[i-1,j-1] \; if \; next \; step \; is \; modifying \; and \; S_1[i] == S_2[j] \quad (3)$$

$$D[i,j] = D[i-1,j-1] + 1 \; if \; next \; step \; is \; modifying \; and \; S_1[i]! = S_2[j] \quad (4)$$

Formula (1),(2),(3),(4), $S_1$ and $S_2$ represent the two sequences compared, respectively, and D[i,j] represents the distance from the first character to the 'i' in $S_1$ and the distance from the first character to the j in $S_2$.

Since the distance from any string to the empty string is the length of the string itself, We have the following formula for initializing D:

$$D[i,0] = i \; and \; D[0,j] = j \quad (5)$$

From (1), (2), (3), (4), (5) we can get the edit distance of the two sequences.

Random Forest: We will use the CDR3 and antigen sequence encoded by N-gram as feature inputs and labels as model outputs to construct a random forest model(RF).

The model construction is shown in the Figure 2:

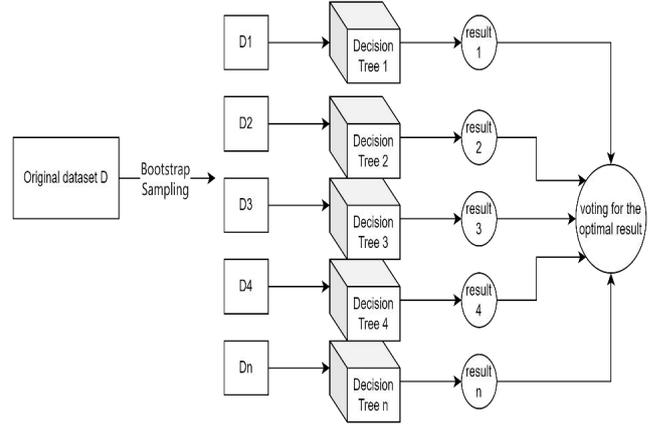

Fig. 2 RF classification principle

For the construction of decision trees, we use the GINI coefficient for evaluation, as shown by the following formula:

$$GINI(D) = 1 - \sum_{i=1}^{k} (P_i)^2 \quad (6)$$

In formula (6), D represents the dataset, k represents the number of categories, and $P_i$ represents the proportion of samples belonging to category i in the dataset.

Fold validation: Regarding model evaluation and comparison, we adopted k-fold validation rather than only through the model's performance in the test set. We divide the data into k subsets, one of which is the validation set, and the other K-1 subset is the training set. The training set is used to train the model, the validation set is used to calculate the performance, and the average performance of the k validation results is calculated. This approach reduces the risk of chance and leads to a more reliable assessment.

IV. DATA DESCRIPTION AND PREPARATION

**Data Description**

VDJdb is a comprehensive antigen-specific TCR sequence database containing manually curated results of published T cell specificity assays[6]. This dataset is widely used in immunomics research, providing researchers with rich information for exploring and analyzing the interaction between immune receptors and antigens and contributing to the development of immunomics.

Based on the detailed explanation of VDJDB[8], we have summarized the meanings of each attribute. Due to the detailed descriptions in the VDJDB documentation, the article will not repeat the meanings of each attribute. For the task of TCR specificity testing, the following 12 attributes are considered valuable: complex.id, cdr3, gene, v.segm, j.segm, antigen.Epitope, antigen.gene, species, vdjdb.Score, mhc. a, mhc. b, mhc.class.

**Data preparation**

Dataset preprocessing: We first need to detect missing values in the database, as shown in Table 1:

Table 1. Missing value detection

| Index | Name | Missing values |
|---|---|---|
| 0 | complex.id | 0 |
| 1 | cdr3 | 0 |
| 2 | gene | 0 |
| 3 | v.segm | 0 |
| 4 | j.segm | 0 |
| 5 | antigen.epitope | 0 |
| 6 | antigen.gene | 0 |
| 7 | species | 0 |
| 8 | vdjdb.score | 0 |
| 9 | mhc.a | 0 |
| 10 | mhc.b | 0 |
| 11 | mhc.class | 0 |

From Table 1, it can be seen that there are all the values in the database for the attributes we need. Next, we conduct Spearman correlation analysis on these 12 attributes, and the results of the correlation analysis can help us screen for attributes with collinearity, thereby reducing the number of data attributes that the model needs to use. Our Spearman correlation analysis results are shown in Figure 3:

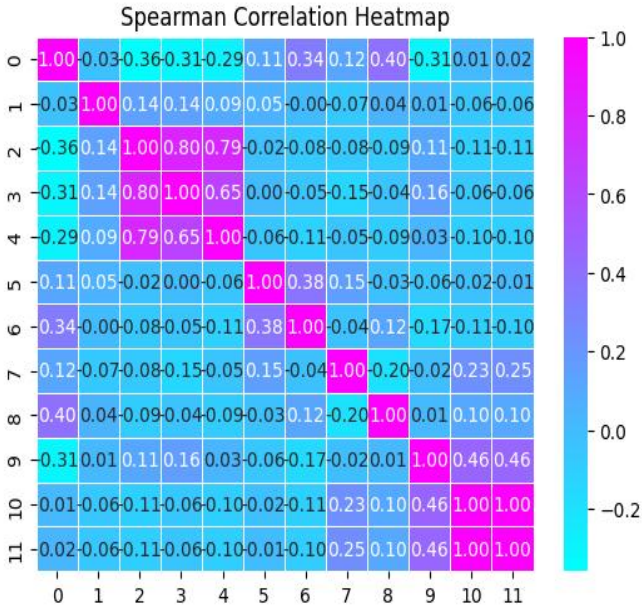

Fig.3 Correlation Analysis

The graph shows that mhc.class and mhc.b are entirely correlated, and j.segm, v.segm, and gene are highly correlated. So, to avoid the problem of high collinearity in the model, we must refrain from using highly correlated data simultaneously.

Next, we counted the data categories of some attributes, as shown in Figure 4:

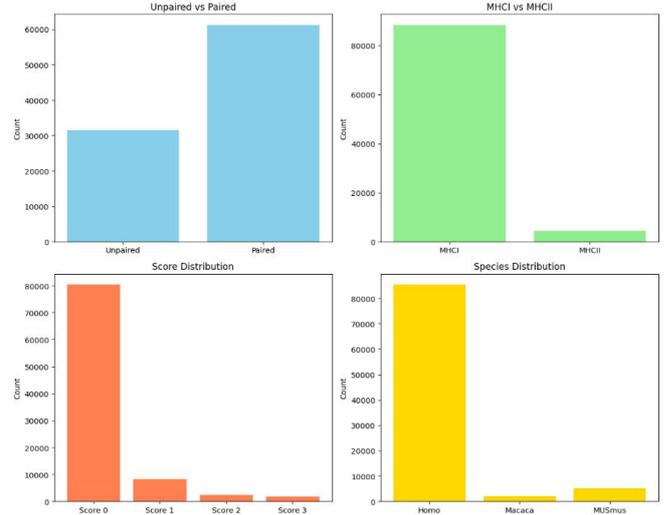

Fig. 4 Data categories for specific attributes

Lu Tianshi's team[9] also focused on the data in the above figure when using deep learning to predict TCR specificity. When processing the data, they filtered out paired data that only selected the MHC category as MHCI and belonged to the human species. In their article, Rudolph, M. G [10] also pointed out that the interaction process between TCR and MHC differs between MHCI and MHCII. Kim, S. M. [11] emphasized the importance of paired data in the article, pointing out that α -and β -chains play a significant role in the function of T cells.

Based on the above analysis, our preprocessing steps are as follows:

Step 1: retain only the complex.id, cdr3, gene, antigen.epitope, species, vdjdb.score, and mhc.class attributes.

Step 2: Filter out data with mhc.class as MHCI.

Step 3: Filter out paired data with complex.id not 0.

Step 4: Filter out the data with species such as HomoSapien.

Step 5: Filter out data with vdjdb.score not being 0.

Step 6: Rearrange the paired datapoint and merge them into the same datapoint(cdr3 with TRA and cdr3 with TRB).

TCR visualization based on PCA and t-SNE： After calculating TRA, TRB, and TRA-TRB separately, we obtained three different distance matrices D1, D2, and $D_3$ , based on edit distance. To speed up the subsequent nonlinear dimensionality reduction, we first use linear dimensionality reduction to reduce the size of the distance matrix. Then, we first use PCA to reduce the distance matrix to 50 dimensions and explain the variance as shown in Table 2:

Table 2. PCA explained variance

| Matrix | $D_1$ | $D_2$ | $D_3$ |
|---|---|---|---|
| Base | TRA | TRB | TRA-TRB |
| variance | 0.96041 | 0.965254 | 0.93969 |

After using PCA to reduce dimensionality to 50 dimensions, it can be seen that most of the information was retained. We do not need to discuss the variance of the individual features; we need to use the reduced matrix for further calculation. Then, we used t-SNE for dimensionality reduction, and the results of dimensionality reduction are shown in the following Figure 5:

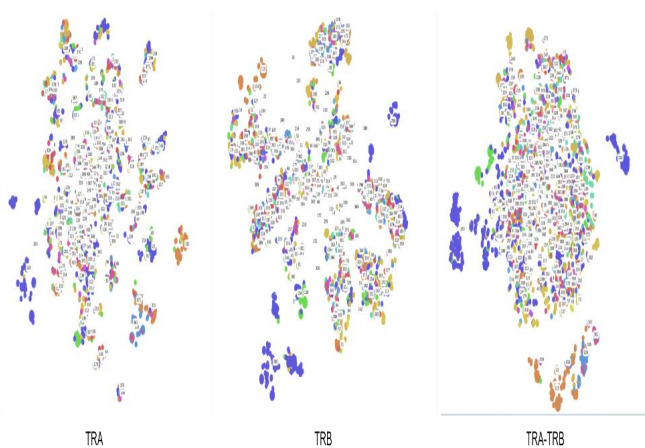

Fig. 5 Visualization after dimensionality reduction

Figure 3 shows that the clusters of the distance matrix $D_2$ based on TRB are more apparent, and the distances of TCRs of similar antigens are closer. The clusters based on TRA and TRA-TRB distance matrix $D_1$ and $D3$ could be more prominent, and most of the TCRS are clustered in one cluster.

Clustering TCR based on editing distance: Through Figure 5, directly using the distance matrix will not be successful because TCRS of different antigens are also easily clustered into one class. We used K-means and DBSCAN to verify our conjecture. We use NMI and ARI to evaluate the performance of clustering algorithms. The formula of NMI and ARI is shown in (7) and (8):

$$NMI(C, T) = 2 * I(C, T)/H(C) + H(T) \quad (7)$$

In formula (7), C represents the clustering result, T represents the distribution of the real labels, I (C, T) represents the mutual information of the clustering result and the distribution of the real label, H(C), H(T) represents their entropy.

$$ARI(C, T) = count(C, T)/max(C, T) \quad (8)$$

In formula (8), count(C, T) represents the number of pairs of samples in C and T that are simultaneously assigned to the same cluster, and max(C, T) represents the maximum possible given the total number of samples, and the size of each cluster.

We directly use clustering algorithms to cluster the distance matrix and obtain clustering results, as shown in Table 3 and Table 4:

Table 3. Cluster performance based on k-means

| Matrix | $D_1$ | $D_2$ | $D_3$ |
|---|---|---|---|
| Base | TRA | TRB | TRA-TRB |
| NMI | 0.01507 | 0.02621 | 0.10711 |
| ARI | 0.112103 | 0.113624 | 0.022367 |

As can be seen from the effect of k-means in Table 3, the effect of the clustering algorithm that specifies the number of clustering centers in advance could be better. We try to use DBSCAN without specifying the number of cluster centers in advance, and the clustering effect is shown in Table 4 :

Table 4. Cluster performance based on DBSCAN

| Matrix | $D_1$ | $D_2$ | $D_3$ |
|---|---|---|---|
| Base | TRA | TRB | TRA-TRB |
| NMI | 0.41187 | 0.405358 | 0.352910 |
| ARI | 0.01912 | 0.014908 | 0.00949 |

Table 3 and Table 4 show that directly using clustering algorithms has poor clustering performance on the entire distance matrix. This is because TCRs with specificity for similar antigens will be grouped into the same cluster, and there are many such TCRs in the database. If clustering is only performed on TCRs with dissimilar antigens, the effect will be relatively ideal, as shown in Figure 6:

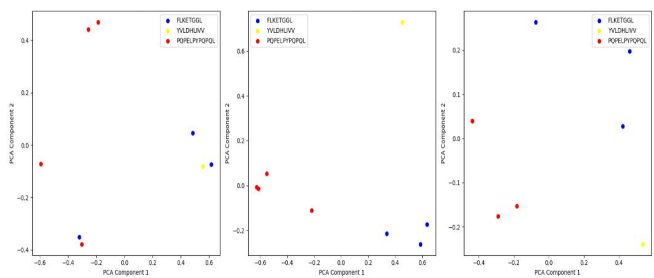

Fig. 6 Dimensionality reduction display of minority points with dissimilar antigens

The clustering effect will be excellent for a small number of TCRs with dissimilar antigens. Still, it is not appropriate for databases with many TCRs to directly use clustering to predict specificity. However, we can still conclude that TCRs with close editing distances have similar antigens.

We then compute the TRB-based edit distance matrix for the test data, using the antigen of the training data with the closest edit distance as the predicted antigen. We calculated an average accuracy of 66.6%. We found that using the antigen from the nearest training data effectively predicts the antigen. However, a significant error remains—thirty percent of the data needed to predict the antigen correctly. We then calculated whether the predicted antigen was present in the first five nearest training data with a 71 percent probability. The calculation process is shown in Figure 7:

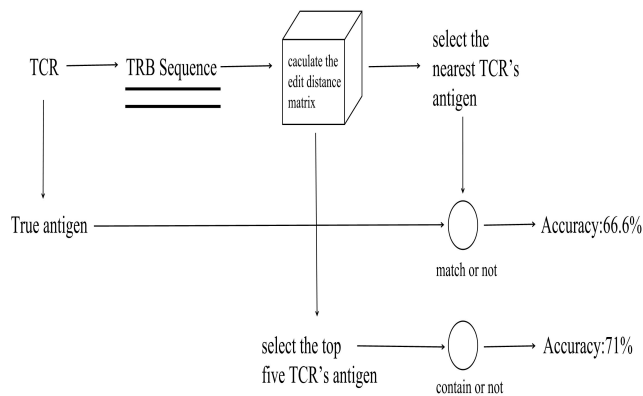

Fig.7 Calculation process

We need a more accurate and efficient way of calculating. So the idea of the first part of our model is to generate the target antigen through a distance matrix: calculate the edit distance matrix of the TRB between the target TCR and other TCRs in the training database, calculate the closest training data, and use our subsequent classifier for further more accurate discrimination.

Construction of Random Forest: For encoding cdr3 and antigen sequences, we adopt the N-gram encoding method, which has the advantage of capturing the local structural information of cdr3 and antigen well while retaining the length distribution information of cdr3. The formula is as follows:

$$\text{Encode(CDR3)} = \text{Ngram(CDR3, N, feature)} \quad (9)$$

We can adjust our encoding method by searching for the optimal N and feature size according to the formula.

In addition to the N-gram encoding method, we also discussed the encoding methods of one-hot representation(10), improved one-hot representation(11), and word bags representation(12) and compared the performance of each encoding method in the experiment. Their formula is shown below:

$$\text{Encode(CDR3)} = \text{padding}(\text{onehotmapping(CDR3)}) \quad (10)$$

$$\text{Encode(CDR3)} = \text{padding}(\text{onehotmapping(CDR3)}) + \text{length(CDR3)} \quad (11)$$

$$\text{Encode(CDR3)} = \text{Wordbag(CDR3)} \quad (12)$$

In the discussion section, we demonstrated the excellent performance of N-gram encoding by comparing other encoding methods.

We use data with vdjdb.score 0 as negative class data (0 labels) and data with vdjdb.score 1, 2, and 3 as positive class data (1 label). Due to the large number of negative class data compared to positive class data, to reduce model bias caused by class imbalance, we adopt a random sampling method to sample negative class data, randomly selecting the same amount of negative class data as positive class data as training data.

Next, we used N-gram models to encode the CDR3 and antigen receptor sequences, respectively. The hyperparameters of the model are shown in Table 5:

Table 5. Hyperparameter settings

| parameter | analyzer | max features | ngram_range | lowercase |
|---|---|---|---|---|
| value | char | 2000 | (5,5) | False |

Selecting how many evaluators (number of decision trees) for a random forest is a vital hyperparameter. We use k-fold validation to calculate the average accuracy of samples for different numbers of random forest models, as shown in Figure 8:

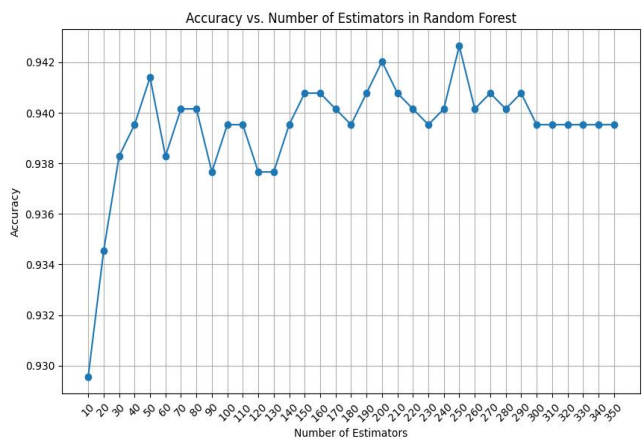

Fig.8 Accuracy vs. Number of Estimators in Random Forest

From the graph, we can intuitively see that the model achieves the best accuracy when the number of decision trees is 250. So, we have chosen 250 evaluators, and our choices for other hyperparameters and k-fold validation settings are shown in Table 6:

Table 6. hyperparameters settings

| parameter | setting |
|---|---|
| max_depth | [None, 5, 10, 15] |
| max_features | ['auto', 'sqrt', 'log2'] |
| min_samples_split | [2, 5, 10] |
| cv | 5 |
| scoring | accuracy |

The final hyperparameter design determined by grid search is shown in the Table 7:

Table 7. Final hyperparameter design

| parameter | setting |
|---|---|
| max_depth | None |
| max_features | log2 |
| min_samples_split | 2 |
| n_estimators | 250 |

Based on the above parameter Settings, we visualize the structure of a decision tree in RF as Figure 9:

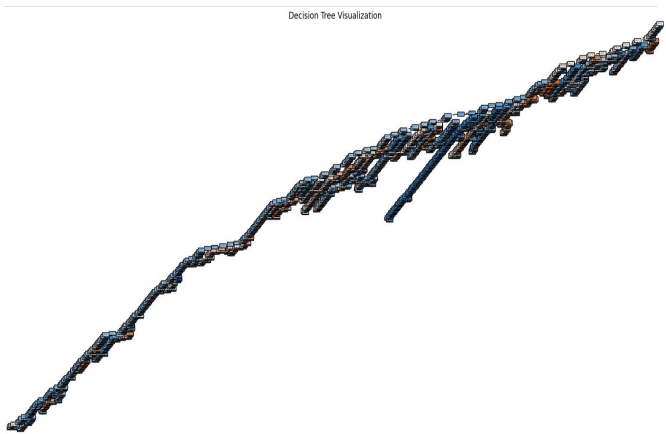

Fig. 9 Structure of a decision tree

## V. RESULTS AND DISCUSSIONS

**Result**

Based on our test set, we compared the results of our generative model with the proper antigen of TCR. The confusion matrix of the model results is shown in the Figure 10:

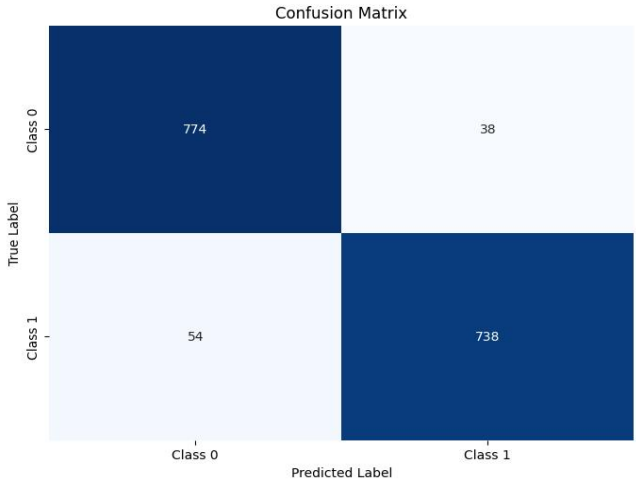

Fig.10 Confusion matrix

As can be seen from the confusion matrix, only a minimal number of TCRS and antigens were mispredicted. The TCR-antigen pairs were correctly classified. Meanwhile, we calculated the common performance indicators in Table 8:

Table 8. Model performance

| Indicator | Accuracy | Recall | Precision | F1 score |
|---|---|---|---|---|
| value | 0.9426 | 0.9318 | 0.9510 | 0.9413 |

The conventional evaluation indicators of the model perform very well. Meanwhile, we plotted the ROC curve of the model as shown in Figure 11:

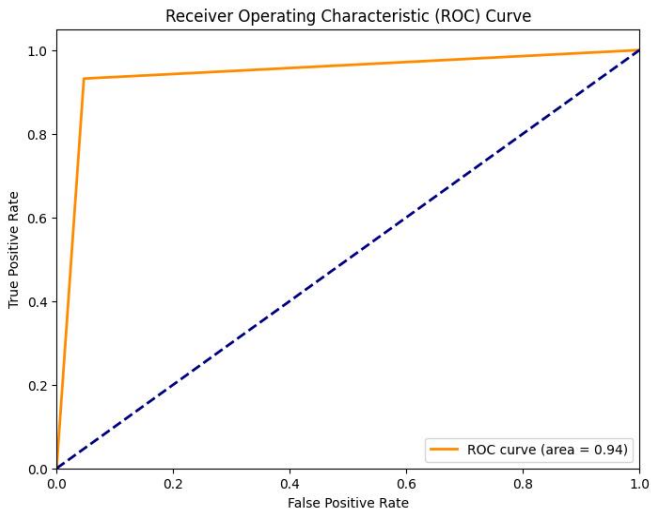

Fig.11 ROC Curve

From the above model results, adding a classifier for discrimination is quite effective compared to antigen prediction using a distance matrix.

The disadvantage of classifiers is that we need to obtain both the antigen sequence and the TCR cdr3 sequence to make predictions, and the application scenarios are relatively small. Our model can better adapt to more application scenarios.

**Discussion**

Next, we will discuss other methods to compare the performance of different models. We calculated the accuracy of other classifier models and encoding methods through k-fold validation (k=5), as shown in Figure 12:

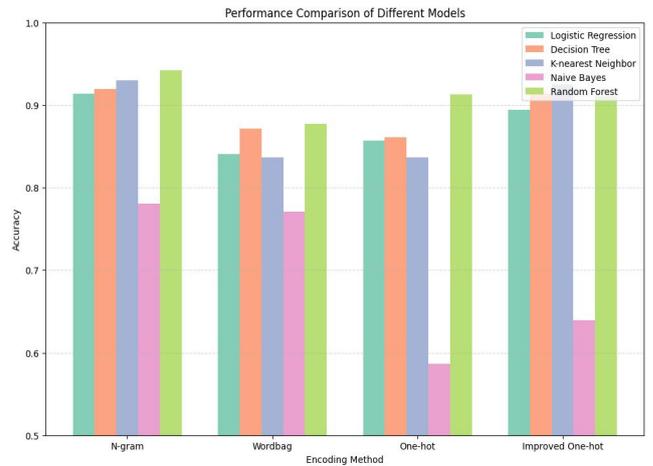

Fig.12 Performance comparison

From the graph, we can see that N-gram+RF is the best-performing classifier.

At the same time, we also constructed two deep learning models, LSTM and CNN, for training. The accuracy of the test set is shown in Figure 13:

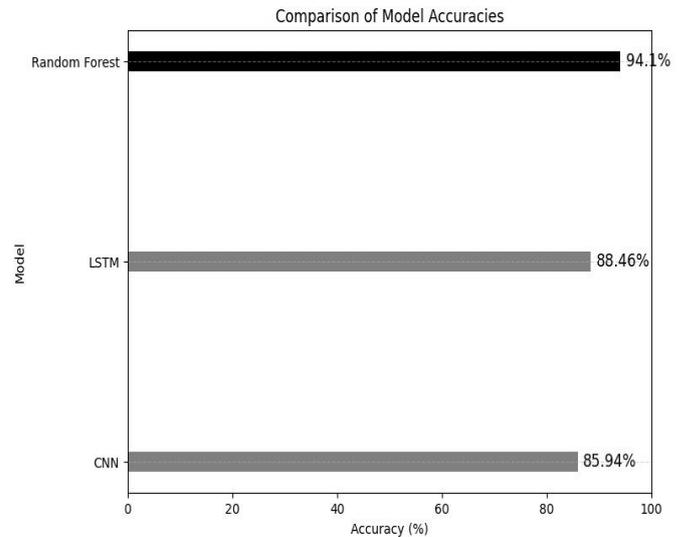

Fig.13 Performance of deep learning model

Even though we adjusted the number of parameters, model structure, regularization techniques, and deep learning multiple times, they did not achieve superior performance than RF on this database.

Table 9. Different algorithms' performance

|  | N-gram | Word-bag | One-hot | Im-one-hot |
|---|---|---|---|---|
| Logistic Regression | 0.91396 | 0.84068 | 0.85666 | 0.8946384 |
| Decision Tree | 0.919576 | 0.871323 | 0.86087 | 0.913341 |
| K-nearest Neighbor | 0.930174 | 0.83700 | 0.83642 | 0.925187 |
| Naive Bayes | 0.78054 | 0.77083 | 0.586846 | 0.639027 |
| Random Forest | 0.942019 | 0.877450 | 0.91317 | 0.93017 |
| CNN | 0.8594 | null | null | null |
| LSTM | 0.8846 | null | null | null |

## VI. FURTHER WORK AND IMPROVEMENT

Enhancement of Datasets: For the task of TCR specificity detection, there are too few high-quality paired data with high reliability in the database (only a few hundred high-quality paired data with score=3). So, we need to enhance our dataset. Based on DeepImmuno-GAN [12], we can also improve the dataset through machine learning methods (GAN) or data from multiple databases. After enhancing the dataset, deep learning models may perform surprisingly.

Improvement of antigen generation algorithm: Even though our algorithm for generating antigens based on edit distance has acceptable performance, for some TCRs, the algorithm efficiency is low due to excessive search times. We can use lower time complexity and more reliable distance algorithms or other generation algorithms to form the first part of the model.

## VII. CONCLUSION

TCR specificity prediction technology has broad prospects in many fields, such as cancer immunotherapy and autoimmune disease treatment. Compared with experimental methods, the computational methods that use algorithms to achieve TCR-specific prediction have advantages like high efficiency, economy, and repeatability. Inspired by previous work, we established a TCR generative specificity detection framework to efficiently screen out TCRs and corresponding antigens through antigen selector and TCR classifier based on random forest. A series of comparisons and analyses indicate that our model outperforms ordinary algorithms.

However, our model still needs to be improved and faces some challenges. First, scarcity and quality of data and high redundancy of the available data still exist (Montemurro et al., 2022b). On the one hand, relevant data collection work is required to obtain richer and higher-quality available data. On the other hand, it is feasible to enhance the dataset using machine learning methods (GAN) or data from multiple databases. Furthermore, the search time may be too long for some TCRs. If we have more time, we will optimize our model's algorithm for selecting target antigens.

REFERENCES

[1] World Health Organization: WHO. (2024, February 1). Global cancer burden growing, amidst mounting need for services.WHO. https://www.who.int/news/item/01-02-2024-global-cancer-burden-growing--amidst-mount ing-need-for-services

[2] Schuster, M., Nechansky, A., & Kircheis, R. (2006). Cancer immunotherapy. Biotechnology Journal, 1(2), 138–147. https://doi.org/10.1002/biot.200500044

[3] Esfahani, K., Roudaia, L., Buhlaiga, N., Del Rincón, S. V., Papneja, N., & Miller, W. H. (2020). A review of cancer immunotherapy: From the past, to the present, to the future. Current Oncology, 27(12), 87–97. https://doi.org/10.3747/co.27.5223

[4] Zhao, M., Xu, X. S., Yang, Y., & Yuan, M. (2023). GGNPTCR: a generative graph structure neural network for predicting immunogenic peptides for T-cell immune response. Journal of Chemical Information and Modeling, 63(23), 7557−7567. https://doi.org/10.1021/acs.jcim.3c01293

[5] Mariuzza, R. A., Agnihotri, P., & Orban, J. (2020). The structural basis of T-cell receptor (TCR) activation: An enduring enigma. Journal of Biological Chemistry/ the Journal of Biological Chemistry, 295(4), 914−925. https://doi.org/10.1016/s0021-9258(17)49904-2

[6] Shugay, M., Bagaev, D. V., Zvyagin, I. V., Vroomans, R. M. A., Crawford, J. C., Dolton, G., Komech, E. A., Sycheva, A. L., Koneva, A. E., Egorov, E. S., Елисеев, A. B., Van Dyk, E., Dash, P., Attaf, M., Rius, C., Ladell, K., McLaren, J. E., Matthews, K., Clemens, E. B., . . . Chudakov, D. M. (2017). VDJdb: a curated database of T-cell receptor sequences with known antigen specificity. Nucleic Acids Research, 46(D1), D419–D427. https://doi.org/10.1093/nar/gkx760

[7] Vujović, M., Degn, K., Marin, F. I., Schaap‐Johansen, A., Chain, B., Andresen, T. L., Kaplinsky, J., & Marcatili, P. (2020). T cell receptor sequence clustering and antigen specificity. Computational and Structural Biotechnology Journal, 18, 2166 − 2173. https://doi.org/10.1016/j.csbj.2020.06.041

[8] Antigenomics. (2023, June). vdjdb-db. GitHub. https://github.com/antigenomics/vdjdb-db/blob/2023-06-01/README.md.

[9] Lu, T., Zhang, Z., Zhu, J., Wang, Y., Jiang, P., Xiao, X., Bernatchez, C., Heymach, J. V., Gibbons, D. L., Wang, J., Xu, L., Reuben, A., & Wang, T. (2021). Deep learning-based prediction of the T cell receptor−antigen binding specificity. Nature Machine Intelligence, 3(10), 864−875. https://doi.org/10.1038/s42256-021-00383-2

[10] Rudolph, M., & Wilson, I. A. (2002b). The specificity of TCR/pMHC interaction. Current Opinion in Immunology, 14(1), 52 − 65. https://doi.org/10.1016/s0952-7915(01)00298-9

[11] Kim, S., Bhonsle, L., Besgen, P., Nickel, J., Backes, A., Held, K., Vollmer, S., Dornmair, K., & Prinz, J. C. (2012). Analysis of the Paired TCR $\alpha$- and $\beta$-chains of Single Human T Cells. PloS One, 7(5), e37338. https://doi.org/10.1371/journal.pone.0037338

[12] Li, G., Iyer, B., Prasath, V. B. S., Ni, Y., & Salomonis, N. (2021b). DeepImmuno: deep learning-empowered prediction and generation of immunogenic peptides for T-cell immunity. Briefings in Bioinformatics, 22(6). https://doi.org/10.1093/bib/bbab160